\documentclass[aps,prd,twocolumn,superscriptaddress,showpacs,amsmath,amssymb]{revtex4}

\usepackage{graphicx}% Include figure files
\usepackage{dcolumn}% Align table columns on decimal point
\usepackage{lineno}
\usepackage{setspace}

\usepackage{setspace}
\usepackage{subfigure}
\usepackage{slashbox}
\usepackage{epsf}
\usepackage{epsfig}

\begin{document}

\newcommand{\mtop}{\mbox{$M_{\rm top}$}}
\newcommand{\ttbar}{\mbox{$t\bar{t}$}}
\newcommand{\thetastar}{\mbox{$\theta^*$}}
\newcommand{\costhetastar}{\mbox{$\cos\thetastar$}}
\newcommand{\Mlb}{\mbox{$M_{lb}$}}
\newcommand{\Et}{\mbox{$E_T$}}
\newcommand{\Pt}{\mbox{$p_T$}}
\newcommand{\Flong}{\mbox{$F_0$}}
\newcommand{\Fplus}{\mbox{$F_+$}}
\newcommand{\met}{\mbox{$\protect \raisebox{0.3ex}{$\not$}\Et$}}

% You should use BibTeX and revtex.bst for references
\bibliographystyle{revtex}

\title{\boldmath         Measurement of the  Top-Quark Mass in the All-Hadronic Channel using the full CDF data set}

%% Author List
% Last update: $Date: 2014/07/09 14:55:49 $
\affiliation{Institute of Physics, Academia Sinica, Taipei, Taiwan 11529, Republic of China}
\affiliation{Argonne National Laboratory, Argonne, Illinois 60439, USA}
\affiliation{University of Athens, 157 71 Athens, Greece}
\affiliation{Institut de Fisica d'Altes Energies, ICREA, Universitat Autonoma de Barcelona, E-08193, Bellaterra (Barcelona), Spain}
\affiliation{Baylor University, Waco, Texas 76798, USA}
\affiliation{Istituto Nazionale di Fisica Nucleare Bologna, \ensuremath{^{ii}}University of Bologna, I-40127 Bologna, Italy}
\affiliation{University of California, Davis, Davis, California 95616, USA}
\affiliation{University of California, Los Angeles, Los Angeles, California 90024, USA}
\affiliation{Instituto de Fisica de Cantabria, CSIC-University of Cantabria, 39005 Santander, Spain}
\affiliation{Carnegie Mellon University, Pittsburgh, Pennsylvania 15213, USA}
\affiliation{Enrico Fermi Institute, University of Chicago, Chicago, Illinois 60637, USA}
\affiliation{Comenius University, 842 48 Bratislava, Slovakia; Institute of Experimental Physics, 040 01 Kosice, Slovakia}
\affiliation{Joint Institute for Nuclear Research, RU-141980 Dubna, Russia}
\affiliation{Duke University, Durham, North Carolina 27708, USA}
\affiliation{Fermi National Accelerator Laboratory, Batavia, Illinois 60510, USA}
\affiliation{University of Florida, Gainesville, Florida 32611, USA}
\affiliation{Laboratori Nazionali di Frascati, Istituto Nazionale di Fisica Nucleare, I-00044 Frascati, Italy}
\affiliation{University of Geneva, CH-1211 Geneva 4, Switzerland}
\affiliation{Glasgow University, Glasgow G12 8QQ, United Kingdom}
\affiliation{Harvard University, Cambridge, Massachusetts 02138, USA}
\affiliation{Division of High Energy Physics, Department of Physics, University of Helsinki, FIN-00014, Helsinki, Finland; Helsinki Institute of Physics, FIN-00014, Helsinki, Finland}
\affiliation{University of Illinois, Urbana, Illinois 61801, USA}
\affiliation{The Johns Hopkins University, Baltimore, Maryland 21218, USA}
\affiliation{Institut f\"{u}r Experimentelle Kernphysik, Karlsruhe Institute of Technology, D-76131 Karlsruhe, Germany}
\affiliation{Center for High Energy Physics: Kyungpook National University, Daegu 702-701, Korea; Seoul National University, Seoul 151-742, Korea; Sungkyunkwan University, Suwon 440-746, Korea; Korea Institute of Science and Technology Information, Daejeon 305-806, Korea; Chonnam National University, Gwangju 500-757, Korea; Chonbuk National University, Jeonju 561-756, Korea; Ewha Womans University, Seoul, 120-750, Korea}
\affiliation{Ernest Orlando Lawrence Berkeley National Laboratory, Berkeley, California 94720, USA}
\affiliation{University of Liverpool, Liverpool L69 7ZE, United Kingdom}
\affiliation{University College London, London WC1E 6BT, United Kingdom}
\affiliation{Centro de Investigaciones Energeticas Medioambientales y Tecnologicas, E-28040 Madrid, Spain}
\affiliation{Massachusetts Institute of Technology, Cambridge, Massachusetts 02139, USA}
\affiliation{University of Michigan, Ann Arbor, Michigan 48109, USA}
\affiliation{Michigan State University, East Lansing, Michigan 48824, USA}
\affiliation{Institution for Theoretical and Experimental Physics, ITEP, Moscow 117259, Russia}
\affiliation{University of New Mexico, Albuquerque, New Mexico 87131, USA}
\affiliation{The Ohio State University, Columbus, Ohio 43210, USA}
\affiliation{Okayama University, Okayama 700-8530, Japan}
\affiliation{Osaka City University, Osaka 558-8585, Japan}
\affiliation{University of Oxford, Oxford OX1 3RH, United Kingdom}
\affiliation{Istituto Nazionale di Fisica Nucleare, Sezione di Padova, \ensuremath{^{jj}}University of Padova, I-35131 Padova, Italy}
\affiliation{University of Pennsylvania, Philadelphia, Pennsylvania 19104, USA}
\affiliation{Istituto Nazionale di Fisica Nucleare Pisa, \ensuremath{^{kk}}University of Pisa, \ensuremath{^{ll}}University of Siena, \ensuremath{^{mm}}Scuola Normale Superiore, I-56127 Pisa, Italy, \ensuremath{^{nn}}INFN Pavia, I-27100 Pavia, Italy, \ensuremath{^{oo}}University of Pavia, I-27100 Pavia, Italy}
\affiliation{University of Pittsburgh, Pittsburgh, Pennsylvania 15260, USA}
\affiliation{Purdue University, West Lafayette, Indiana 47907, USA}
\affiliation{University of Rochester, Rochester, New York 14627, USA}
\affiliation{The Rockefeller University, New York, New York 10065, USA}
\affiliation{Istituto Nazionale di Fisica Nucleare, Sezione di Roma 1, \ensuremath{^{pp}}Sapienza Universit\`{a} di Roma, I-00185 Roma, Italy}
\affiliation{Mitchell Institute for Fundamental Physics and Astronomy, Texas A\&M University, College Station, Texas 77843, USA}
\affiliation{Istituto Nazionale di Fisica Nucleare Trieste, \ensuremath{^{qq}}Gruppo Collegato di Udine, \ensuremath{^{rr}}University of Udine, I-33100 Udine, Italy, \ensuremath{^{ss}}University of Trieste, I-34127 Trieste, Italy}
\affiliation{University of Tsukuba, Tsukuba, Ibaraki 305, Japan}
\affiliation{Tufts University, Medford, Massachusetts 02155, USA}
\affiliation{University of Virginia, Charlottesville, Virginia 22906, USA}
\affiliation{Waseda University, Tokyo 169, Japan}
\affiliation{Wayne State University, Detroit, Michigan 48201, USA}
\affiliation{University of Wisconsin, Madison, Wisconsin 53706, USA}
\affiliation{Yale University, New Haven, Connecticut 06520, USA}

\author{T.~Aaltonen}
\affiliation{Division of High Energy Physics, Department of Physics, University of Helsinki, FIN-00014, Helsinki, Finland; Helsinki Institute of Physics, FIN-00014, Helsinki, Finland}
\author{S.~Amerio\ensuremath{^{jj}}}
\affiliation{Istituto Nazionale di Fisica Nucleare, Sezione di Padova, \ensuremath{^{jj}}University of Padova, I-35131 Padova, Italy}
\author{D.~Amidei}
\affiliation{University of Michigan, Ann Arbor, Michigan 48109, USA}
\author{A.~Anastassov\ensuremath{^{v}}}
\affiliation{Fermi National Accelerator Laboratory, Batavia, Illinois 60510, USA}
\author{A.~Annovi}
\affiliation{Laboratori Nazionali di Frascati, Istituto Nazionale di Fisica Nucleare, I-00044 Frascati, Italy}
\author{J.~Antos}
\affiliation{Comenius University, 842 48 Bratislava, Slovakia; Institute of Experimental Physics, 040 01 Kosice, Slovakia}
\author{G.~Apollinari}
\affiliation{Fermi National Accelerator Laboratory, Batavia, Illinois 60510, USA}
\author{J.A.~Appel}
\affiliation{Fermi National Accelerator Laboratory, Batavia, Illinois 60510, USA}
\author{T.~Arisawa}
\affiliation{Waseda University, Tokyo 169, Japan}
\author{A.~Artikov}
\affiliation{Joint Institute for Nuclear Research, RU-141980 Dubna, Russia}
% ...

\collaboration{CDF Collaboration}
\altaffiliation[With visitors from]{
\ensuremath{^{a}}University of British Columbia, Vancouver, BC V6T 1Z1, Canada,
\ensuremath{^{b}}Istituto Nazionale di Fisica Nucleare, Sezione di Cagliari, 09042 Monserrato (Cagliari), Italy,
\ensuremath{^{c}}University of California Irvine, Irvine, CA 92697, USA,
\ensuremath{^{d}}Institute of Physics, Academy of Sciences of the Czech Republic, 182~21, Czech Republic,
\ensuremath{^{e}}CERN, CH-1211 Geneva, Switzerland,
\ensuremath{^{f}}Cornell University, Ithaca, NY 14853, USA,
\ensuremath{^{g}}University of Cyprus, Nicosia CY-1678, Cyprus,
\ensuremath{^{h}}Office of Science, U.S. Department of Energy, Washington, DC 20585, USA,
\ensuremath{^{i}}University College Dublin, Dublin 4, Ireland,
\ensuremath{^{j}}ETH, 8092 Z\"{u}rich, Switzerland,
\ensuremath{^{k}}University of Fukui, Fukui City, Fukui Prefecture, Japan 910-0017,
\ensuremath{^{l}}Universidad Iberoamericana, Lomas de Santa Fe, M\'{e}xico, C.P. 01219, Distrito Federal,
\ensuremath{^{m}}University of Iowa, Iowa City, IA 52242, USA,
\ensuremath{^{n}}Kinki University, Higashi-Osaka City, Japan 577-8502,
\ensuremath{^{o}}Kansas State University, Manhattan, KS 66506, USA,
\ensuremath{^{p}}Brookhaven National Laboratory, Upton, NY 11973, USA,
\ensuremath{^{q}}Queen Mary, University of London, London, E1 4NS, United Kingdom,
\ensuremath{^{r}}University of Melbourne, Victoria 3010, Australia,
\ensuremath{^{s}}Muons, Inc., Batavia, IL 60510, USA,
\ensuremath{^{t}}Nagasaki Institute of Applied Science, Nagasaki 851-0193, Japan,
\ensuremath{^{u}}National Research Nuclear University, Moscow 115409, Russia,
\ensuremath{^{v}}Northwestern University, Evanston, IL 60208, USA,
\ensuremath{^{w}}University of Notre Dame, Notre Dame, IN 46556, USA,
\ensuremath{^{x}}Universidad de Oviedo, E-33007 Oviedo, Spain,
\ensuremath{^{y}}CNRS-IN2P3, Paris, F-75205 France,
\ensuremath{^{z}}Universidad Tecnica Federico Santa Maria, 110v Valparaiso, Chile,
\ensuremath{^{aa}}The University of Jordan, Amman 11942, Jordan,
\ensuremath{^{bb}}Universite catholique de Louvain, 1348 Louvain-La-Neuve, Belgium,
\ensuremath{^{cc}}University of Z\"{u}rich, 8006 Z\"{u}rich, Switzerland,
\ensuremath{^{dd}}Massachusetts General Hospital, Boston, MA 02114 USA,
\ensuremath{^{ee}}Harvard Medical School, Boston, MA 02114 USA,
\ensuremath{^{ff}}Hampton University, Hampton, VA 23668, USA,
\ensuremath{^{gg}}Los Alamos National Laboratory, Los Alamos, NM 87544, USA,
\ensuremath{^{hh}}Universit\`{a} degli Studi di Napoli Federico I, I-80138 Napoli, Italy
}
\noaffiliation
% Last update: $Date: 2014/07/09 14:55:49 $

\begin{abstract}
The top-quark mass $M_{\rm top}$ is measured using top quark-antiquark  pairs produced in 
proton-antiproton  collisions at a center-of-mass energy of  $1.96$ TeV and 
decaying into a fully hadronic final state. The full data set collected with the CDF\,II 
detector at the Fermilab Tevatron Collider, corresponding to an integrated luminosity of $9.3$ fb$^{-1}$, is used.
Events are selected that have  six to eight jets, at least one of which is identified as having originated from a $b$ quark. 
In addition, a multivariate algorithm, containing multiple kinematic variables as inputs, is used to discriminate signal events from background events due to 
QCD multijet production.
Templates for the reconstructed top-quark mass are combined in a likelihood  fit to measure
 $M_{\rm top}$ with a simultaneous calibration of the jet-energy scale.
A value of
\mbox{$M_{\rm top} = 175.07~\pm 1.19\,(\textnormal{stat}) ~ ^{+1.55}_{-1.58}(\textnormal{syst})$~GeV/$c^2$}
is obtained for the top-quark mass.

\end{abstract}

%% keywords

\maketitle

The mass of the top quark, $M_{\rm top}$, is a fundamental parameter 
of the standard model (SM). 
Furthermore, the measured value of $M_{\rm top}$  is comparable to the mass scale of electroweak-symmetry breaking, suggesting that the top quark may play a special role in this phenomenon, either in the SM or in new physics processes beyond the SM\,\cite{topcolor-ref,ewk-break-ref}.
 After the Higgs-boson discovery by the ATLAS and CMS experiments\,\cite{atlas_higgs,cms_higgs}, precise measurements of $M_{\rm top}$ are critical inputs to global electroweak fits that assess the self-consistency of the SM\,\cite{ewk_fits}, and are crucial for determining the stability of the vacuum\,\cite{vacuum_stability}.

\par
In $p\bar p$ collisions at 1.96\,TeV  center-of-mass energy top quarks are produced predominantly in pairs ($t\bar{t}$),
with each top quark decaying 
into a $W$ boson and a bottom quark with a probability of nearly 100\%\,\cite{pdg}.
For this analysis candidate events are selected in which both $W$ bosons decay to a quark-antiquark pair 
($t\bar{t} \to W^{+} b \, W^{-}\bar{b} \to q_{1}\bar{q}_{2} b \, q_{3} \bar{q}_{4} \bar{b}$). 
This final state, the all-hadronic channel, comprises 46\% of all $t\bar t$ final states, which is larger than the probabilities of all other 
individual $t\bar t$ decay channels.
However, it suffers from large multijet background due to quantum chromodynamics (QCD) production, which exceeds  
$t\bar t$ production by three orders of magnitude. The principal advantage of this analysis channel, though, is that a full kinematic reconstruction of the $t\bar t$ state is possible as there are no undetected particles.  
In this paper, we present a measurement of the top-quark mass using the full data set collected by the CDF experiment in 2002-2011,
  with the same event selection as in Ref.\,\cite{ahplb}.
Apart from the nearly two-fold increase in integrated luminosity, additional improvements come from the use of a new Monte Carlo generator. The simulated  samples used for the $t\bar t$ signal are now produced by {\sc powheg}\,\cite{powheg},  a next-to-leading-order generator in the strong-interaction coupling interfaced with {\sc pythia}\,\cite{pyt} for parton shower evolution and hadronization.
\par
The CDF II detector consists of
high-precision tracking systems for vertex and charged-particle track reconstruction, surrounded by electromagnetic and hadronic calorimeters for energy measurement.
Muon subsystems are located outside the calorimeter for muon detection. A detailed description can be found in Ref.\,\cite{CDFdetector}.   
The data correspond to the full integrated luminosity of $9.3$\,fb$^{-1}$.
Events are selected with a multijet trigger\,\cite{ahprd},  
and retained only if they
have no well-identified energetic  electron or muon.  A jet is identified as a cluster of calorimeter energies contained within a cone of
radius $\Delta R\equiv \sqrt{(\Delta\eta)^2+(\Delta\phi)^2}=0.4$, where $\Delta\eta$ and $\Delta\phi$
are the distances in pseudorapidity\,\cite{coordinate} 
and azimuthal angle between a tower center and the cluster axis. Jet energies are corrected
for a number of effects that bias their measurement\,\cite{jes}.
\par
A total of about $11.4\times 10^6$ events are selected in data having six to eight jets, each with a transverse energy of at least 15 GeV and satisfying a pseudorapidity 
requirement of $\vert\eta\vert\le 2.0$. Events with neutrinos in the final state are suppressed by the requirement that the
 missing transverse energy $\met$\,\cite{coordinate} 
is small with respect to its resolution, and 
satisfies  \mbox{$\met/\sqrt{\sum E_{T}}<3$\,GeV$^\frac{1}{2}$},
where $\sum E_{T}$ is the sum of the transverse energy of all jets.
Of these events, less than 16\,000 are expected to originate from $t\bar{t}$ signal.
The signal purity is improved through
an artificial neural network, which takes as input a set of kinematic and jet-shape variables~\cite{ahprd}. 
The neural network is trained using simulated $t\bar{t}$ events for the signal and the selected candidate events for the multijet background, since 
the fraction of $t\bar t$ events in the candidate sample is still negligible (on the order of 1/700).
The value of the output node $N_{\rm out}$ is used as a discriminant between signal and background.
An additional enhancement of the  signal purity comes from the application of a $b$-tagging algorithm. This analysis uses the {\sc SECVTX} algorithm\,\cite{vertex} 
to identify (``tag'') jets 
that most likely originate from the fragmentation of a $b$ quark, requiring the presence of particle trajectories (tracks) forming reconstructable vertices significantly displaced from the vertex of the $p\bar p$ collision. These vertices need to be found inside the jet cone, and 
jet energy corrections specific to $b$-jets are applied to tagged jets.  Only events with one, two, or three tagged jets are kept,
excluding larger multiplicities to reduce the possible assignments of jets to partons in the event reconstruction.
When three $b$-tagged jets are present, the three possible
assignments with two $b$-tagged jets and one light-flavor jet are considered. 
\par
The dominant backgrounds to the all-hadronic final state comes  from the QCD production of 
heavy-quark pairs ($b\bar b$ and $c\bar c$) and from events with incorrectly tagged jets associated with light quarks or gluons. 
Given the large theoretical uncertainties on the QCD multijet production cross section,  it is preferable to infer the background from the data directly.
The ``tag rate'' is defined as the probability  
of tagging a jet, parametrized in terms of jet $E_T$, number of tracks contained in the jet cone, 
and the number of reconstructed primary vertices in the event. This tag rate is obtained in a background-rich control sample with five jets 
and is used to estimate the probability that a candidate event from background
contains a given number of tagged jets.
Before the $b$-tagging requirement is imposed, a probability
is calculated for each data event that one, two, or three jets could be tagged as $b$-jets.
The sum of these probabilities over all pretagged data events represents the background prediction for the given tag category.
Correction factors are introduced to take into account
correlations among jets due to the presence of multiple $b$ quarks in the same event. 
The procedure, described in detail in Ref.\,\cite{ahprd}, allows the prediction of
the expected amount of background  in the selected samples as well as the distributions of specific measured variables, as discussed later. 
\par
The top-quark mass is measured using a ``template method''\,\cite{2dfit}, 
while simultaneously ({\em in situ}) calibrating the jet-energy scale (JES)
to reduce the associated systematic uncertainty. 
Reference distributions (``templates'') are derived for the signal from  variables sensitive to the true values of
$M_{\rm top}$ and  JES. The  chosen templates correspond to the top-quark mass $m_{t}^{\rm rec}$
and the $W$-boson mass $m_{W}^{\rm rec}$, obtained from a kinematical reconstruction of the final state.
The ${\rm JES}$ is a multiplicative factor that,
applied to the raw energy of a reconstructed jet,
returns a corrected energy that is designed to give the best estimate of the energy of the associated parton.
The uncertainty on the  ${\rm JES}$ value to be applied in simulated events 
results in a large uncertainty on the measurements of $M_{\rm top}$.
A maximum likelihood fit is then performed to find the $M_{\rm top}$ and  JES values
that best match the distributions observed in the data.

In this analysis the applied ${\rm JES}$ is expressed as a function of the dimensionless parameter
$\Delta_{\rm JES}$, which measures the shift $\Delta_{\rm JES}$\,$\cdot$\,$\sigma_c$ with respect the CDF default value.
The latter is based on a combination of instrumental calibration and analysis of data control samples\,\cite{jes}, and
$\sigma_c$ represents here its uncertainty. 

For each selected event,  mass combinations are generated\,\cite{ahprd} assigning in turn
each one of the six highest-$E_{T}$ jets to one of the final-state six quarks.
Then, for each combination, two triplets of jets are associated with the two top quarks, each triplet including a pair of jets (corresponding to the $W$ boson) and a $b$-tagged jet. 
The number of possible combinations is reduced by assigning $b$-tagged jets  to 
$b$ quarks only, resulting in 30, 6, or 18 permutations for events with one, two, or three 
tagged jets, respectively. %%\,\cite{permutations}.

For each combination, a value of  $m_t^{\rm rec}$ is obtained 
through a constrained fit based on the minimization of a $\chi^2$-like function defined as:
\begin{eqnarray}
\chi^{2}_{t} &  =  &  ~\frac{ \big(m_{\mathit jj}^{(1)}-M_W \big)^2\,c^4 }{ \Gamma^2_W } +
                       \frac{ \big(m_{\mathit jj}^{(2)}-M_W \big)^2\,c^4 }{ \Gamma^2_W }  \nonumber \\
             &     & + \frac{ \big(m_{\mathit jjb}^{(1)}-m_t^{\rm rec} \big)^2\,c^4 }{ \Gamma^2_t } +
                       \frac{ \big(m_{\mathit jjb}^{(2)}-m_t^{\rm rec} \big)^2\,c^4}{\Gamma^2_t}   \nonumber \\
             &     & + \sum^6_{i=1} \frac{ \big(p^{\rm fit}_{T,i}-p^{\rm meas}_{T,i} \big)^2}{\sigma^2_i} \nonumber
\end{eqnarray} 
where $m_{\mathit jj}^{(1,2)}$ represent the invariant masses of the two pairs of jets assigned to light-flavor quarks,
while $m_{\mathit jjb}^{(1,2)}$
represent the invariant masses of the triplets including one light-flavor pair and one jet assigned to a $b$ quark. The quantities
$M_{W} = 80.4$\,GeV/$c^2$ and $\Gamma_{W}=2.1$\,GeV are the known measured mass and  width 
of the $W$ boson\,\cite{pdg}, while
$\Gamma_t = 1.5$\,GeV is the estimated natural width of the top quark\,\cite{widt}. 
In the fit, the transverse momenta of the jets $p^{\rm fit}_{T,i}$ are constrained  to their measured values
$p^{\rm meas}_{T,i}$ within their known resolutions $\sigma_i$.
Among all combinations, the one that gives the lowest
value for the minimized $\chi^{2}_{t}$ is selected along with the value of
$m_t^{\rm rec}$ determined by the fit.
An additional fit is introduced for the reconstruction of  $m_{W}^{\rm rec}$, by defining a specific $\chi^{2}$ function, $\chi^{2}_{W}$, where the known $W$-boson mass is replaced by  $m_{W}^{\rm rec}$ and left free to vary. 
Independent distributions for events with exactly one or with two or three tags are built from the $m_{t}^{\rm rec}$ and $m_{W}^{\rm rec}$ values.
\par
Signal templates are formed using simulated events with top-quark masses ranging 
from 167.5 to 177.5\,GeV/$c^2$, in steps of 1.0\,GeV/$c^2$, and with $\Delta_{\rm JES}$ between $-2$ and $+2$, 
in steps of $0.5$.
Background templates are obtained applying the fitting technique to the 
events passing the neural-network selection, but before the $b$-tagging requirement 
(``pretag'' sample)\,\cite{ahprd}. The distributions are formed assigning  to each value 
of $m_t^{\rm rec}$ and $m_W^{\rm rec}$ a weight that is given
by the probability of the event to be from background and to contain tagged jets,
as evaluated from the jet tag rates. The signal presence in the pretag sample is accounted for.
\par  
At this stage, two requirements are imposed on the events: $N_{\rm out} \ge 0.97$ (0.94) and $\chi^2_W \le 2$ (3) for 1 ($\ge 2$) tag events. The events that survive 
these selection criteria comprise the $S_{\rm JES}$ sample, which is used primarily to constrain
the statistical uncertainty on the $\Delta_{\rm JES}$ measurement. A subset of the $S_{\rm JES}$ sample ($S_{M_{\rm top}}$) is obtained by
additionally requiring $\chi^2_t \le 3$ (4) for 1 ($\ge 2$) tag events; the $S_{M_{\rm top}}$ sample is the primary set of events used to
extract the top-quark mass. The $N_{\rm out}$, $\chi^2_W$, $\chi^2_t$ thresholds have been optimized to minimize the statistical
uncertainty on the $M_{\rm top}$ measurement based on simulations. The corresponding signal and background events are
 then used to populate the $m_W^{\rm rec}$ and $m_t^{\rm rec}$ templates for the $S_{\rm JES}$ and $S_{M_{\rm top}}$ subsets, respectively.
\par
Table\,\ref{tab:finalsamples} summarizes the event selection for
events with one tag and with two or three tags, separately.

\begin{table}[h!]

     \caption{Numbers of candidate events  ($N_{\rm obs}$)  and expected signal yield in the two selected samples. For the signal, $M_{\rm top} = 172.5$\,GeV/$c^2$ and $\Delta_{\rm JES} = 0$ are
           used, and  expectations are normalized to the integrated luminosity of the 
           data sample ($9.3$\,fb$^{-1}$) using the theoretical cross section ($7.46$\,pb\,\cite{tt-xsec}). The uncertainty on the signal comes from the uncertainty on the cross section and on the integrated luminosity.
             }

 \begin{center}

\hspace*{-8mm}   \begin{tabular}{l  c  c }

\hline\hline

  \parbox[c][12mm][c]{30mm}{~~~~~~~~Sample}   & $N_{\rm obs}$ &

  \parbox[c][6mm][c]{35mm}
	 { Expected $t\bar{t}$ 
        } \\

\hline

\parbox[c][12.5mm][c]{18mm}{$1$-tag}   

                              \parbox[c][17mm][c]{11mm}
                                   {
                                     $\begin{array}{l}
                                          S_{\rm JES}  \\ \\
                                          S_{M_{\rm top}}
                                      \end{array}$
                                   }                 

                  & $\begin{array}{c}     7890      \\ \\      4130       \end{array}$
                  & $\begin{array}{c} 1886 \pm 150  \\ \\   1270 \pm 101   \end{array}$ \\

\hline

\parbox[c][12.5mm][c]{18mm}{$\geq 2$-tags} 

                     \parbox[c][17mm][c]{11mm}
                            {
                                 $\begin{array}{l}
                                                S_{\rm JES}  \\ \\
                                                S_{M_{\rm top}}
                                            \end{array}$
                             }

                  & $\begin{array}{c}     1758      \\ \\      901      \end{array}$
                  & $\begin{array}{c}  782 \pm 64   \\ \\   514 \pm 42  \end{array}$ \\
 
\hline\hline

       \end{tabular}

     \label{tab:finalsamples}

 \end{center}
\end{table}

\begin{figure}[htbp!]
 \begin{center}
  \begin{tabular}{c}
    
      \epsfig{file=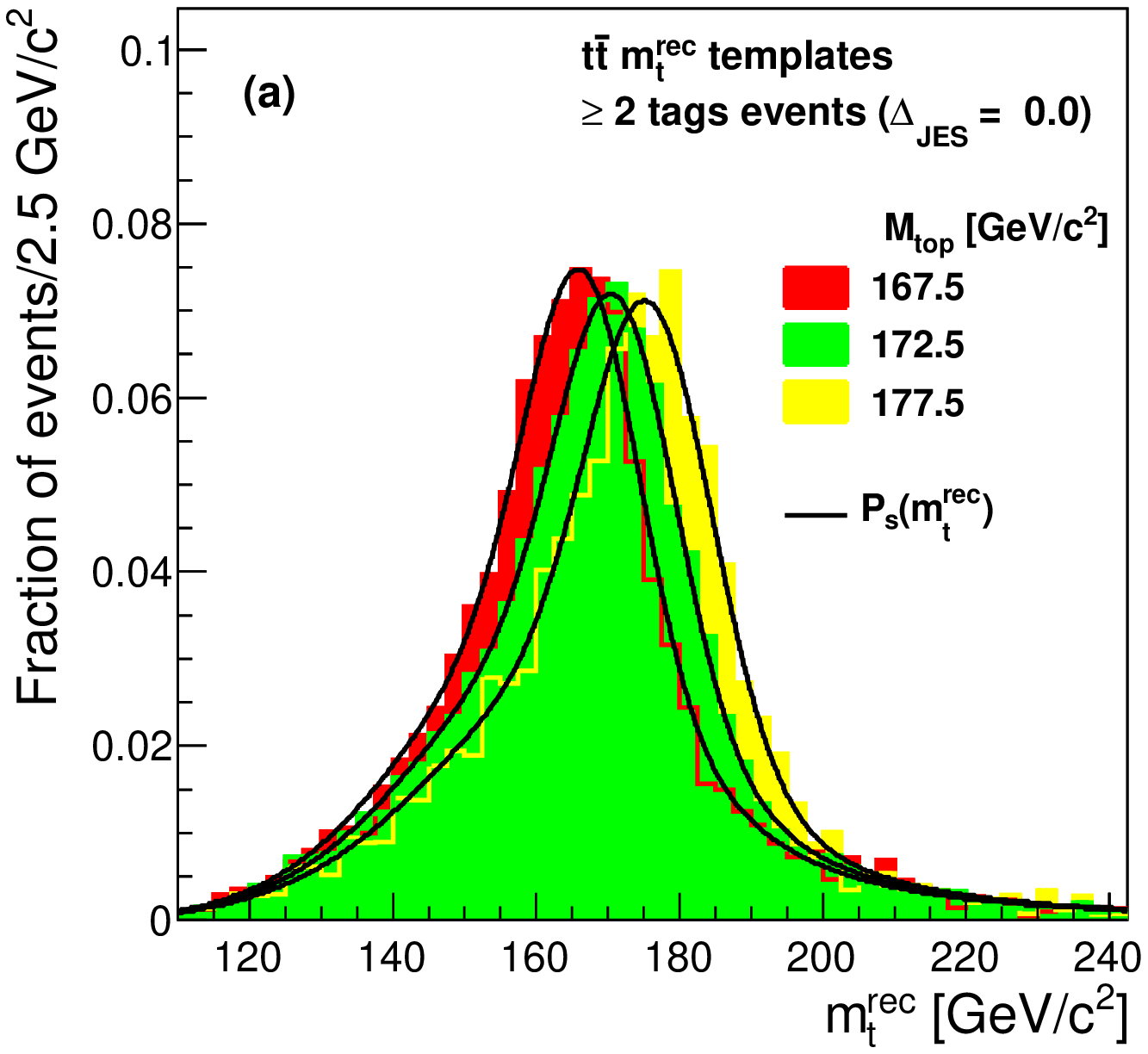,clip=true,bbllx=5pt,bblly=5pt,bburx=480pt,bbury=445pt,width=9cm}  
   \\
      \epsfig{file=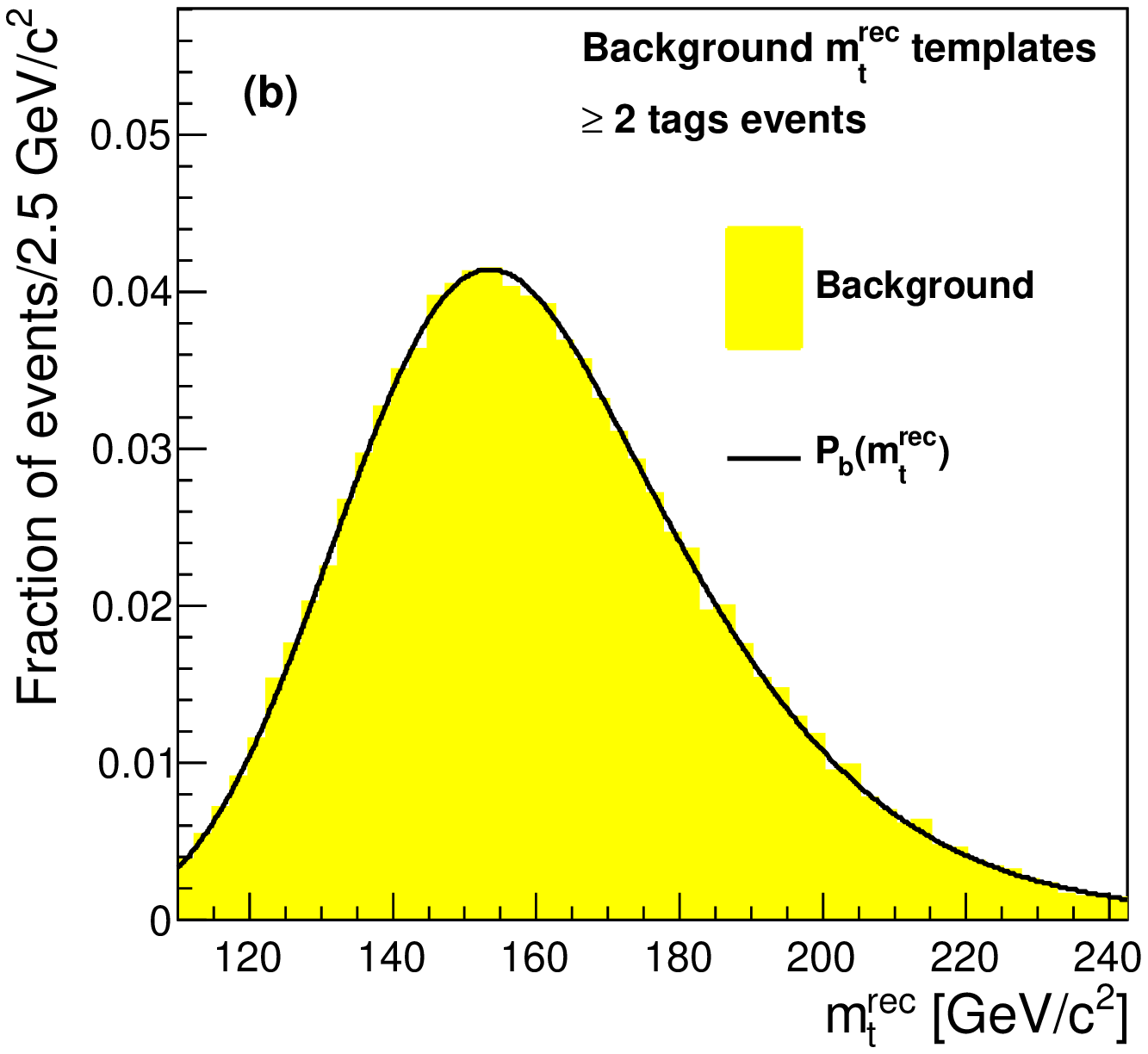,clip=true,bbllx=5pt,bblly=5pt,bburx=480pt,bbury=445pt,width=9cm}    
                          
  \end{tabular}
  \end{center}
   \vspace*{-3mm}
      \caption{
               Templates of $m_{t}^{\rm rec}$ for events with two or three tags and corresponding
               probability density functions superimposed. (a): the signal p.d.f $P_{\rm s}$, 
               for various values of $M_{\rm top}$ and $\Delta_{\rm JES}=0$.   
               (b): the background p.d.f. $P_{\rm b}$.
              }
      \label{fig:templates}
\end{figure}

The measurement of $M_{\rm top}$ and the simultaneous calibration of  $\mathrm{JES}$
are performed by maximizing an
unbinned extended-likelihood function. 
The function, defined in detail in Ref.\,\cite{ahplb}, is divided into three parts,
\begin{center}
$ {\cal L} = {\cal L}_{1\,\mathrm{tag}} \times {\cal L}_{\ge 2\,\mathrm{tags}} \times {\cal L}_{\Delta_{\rm JES,constr}} $,
\end{center}  
where ${\cal L}_{\Delta_{\rm JES,constr}}$ is a Gaussian term constraining the 
$\mathrm{JES}$ to the nominal value (i.e., $\Delta_{\rm JES}$ to $0$) within its uncertainty.
The two terms ${\cal L}_{1\,{\rm tag}}$ and ${\cal L}_{\ge 2 \, \mathrm{tags}}$ are in turn defined as
\begin{eqnarray}
 {\cal L}_{1,\ge 2\, \mathrm{tags}} =       {\cal L}_{\Delta_{\rm JES}} 
                                     \times {\cal L}_{M_{\mathrm{top}}} 
                                     \times {\cal L}_{\mathrm{evts}}\,, \nonumber
\end{eqnarray}
where ${\cal L}_{\mathrm{evts}}$ gives the probability to observe simultaneously the
number of events selected in the $S_{\rm JES}$ and the $S_{M_{\rm top}}$ data samples, given the expected signal and background yields. Unlike the analysis in Ref.\,\cite{ahplb}, the background yields are allowed to vary unconstrained in the fit. 
The two terms $ {\cal L}_{\Delta_{\rm JES}}$ and ${\cal L}_{M_{\mathrm{top}}} $ represent the likelihoods, based on the signal and background templates, to observe the sets of  $m_W^{\mathrm{rec}}$ and $m_t^{\mathrm{rec}}$ values in the two data sets $S_{\rm JES}$ and $S_{M_{{\rm top}}}$. For each signal template, the probability density function (p.d.f.)
 is represented as a sum of Gamma and Gaussian functions, whose parameters are in turn linear functions 
of the fit parameters  $M_{\mathrm{top}}$ and $\Delta_{\rm JES}$. 
In Fig.\,\ref{fig:templates},  examples of signal and background $m_{t}^{\mathrm{rec}}$  templates for the sample with two or three tags are shown,
with the corresponding p.d.f.'s superimposed.

\par
The possible presence of biases in  the values 
returned by the likelihood fit is investigated and taken into account.
Pseudoexperiments (PEs) are performed assuming specific
values for $M_{\rm top}$ and $\Delta_{\rm JES}$
and ``pseudodata'' are extracted 
from the corresponding signal and background templates and subjected to the likelihood maximization procedure.
The results of these PEs are compared to the input values, and 
linear calibration functions are defined to obtain, on 
average, a more accurate estimate of the true values and uncertainties. 
The average shift in top-quark mass due to the calibration is about 200 MeV/$c^2$.
\par
\begin{figure}[htbp!]
\begin{center}
  \begin{tabular}{c}  

      \epsfig{file=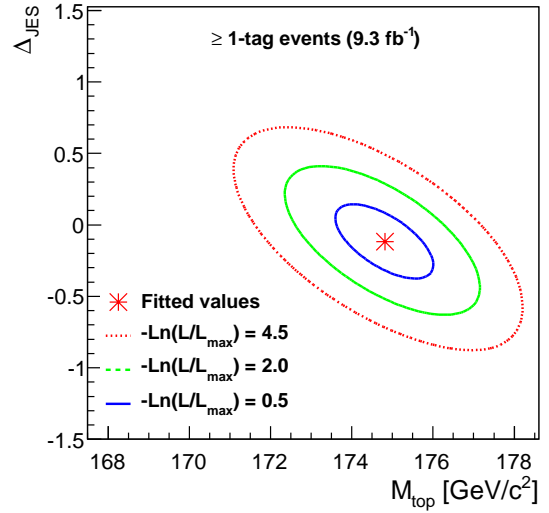,clip=true,bbllx=10pt,bblly=25pt,bburx=382pt,bbury=375pt,width=8cm}

  \end{tabular} 
  \end{center}
  \vspace*{-3mm}
  \caption{ Negative log-likelihood contours for the likelihood fit performed for the 
            $M_{\rm top}$ and  $\Delta_{\rm JES}$ measurement, before calibration, for events with one, two,  or three tags. 
             The minimum is shown along with the contours whose projections correspond to one, two, and three $\sigma$ 
             uncertainties on the $M_{\rm top}$ and $\Delta_{\rm JES}$ measurements.
           }
\label{fig:contour}
\end{figure}

The likelihood fit is applied to the data, and after applying the calibration corrections, the values returned by the fit are:
$$M_{\rm top} =  175.07 \pm 1.19\,(\textnormal{stat}) \pm 0.97\, (\mathrm{JES})\pm 0.41\, (\textnormal{fit}) ~\mathrm{GeV}/c^2,$$
and
$$\Delta_{\rm JES} =  -0.282 \pm 0.255\,(\textnormal{stat}) \pm 0.207 \,(M_{\rm top}) \pm 0.040 \,(\textnormal{fit}),$$
where the fit uncertainties are those arising from the variation in the fitted signal and background yields, to which the additional systematic uncertainties described below will be added in quadrature. The correlation between $M_{\rm top}$ and $\Delta_{\rm JES}$ amounts to $-0.63$. 
The best-fit values of $M_{\rm top}$ and $\Delta_{\rm JES}$ are shown in Fig.\,\ref{fig:contour}, along with  the negative log-likelihood 
contours whose projections correspond to one, two, and three $\sigma$ uncertainties on the values 
of $M_{\rm top}$ and $\Delta_{\rm JES}$. 
The fit returns, for the  $S_{M_{{\rm top}}}$ sample, a signal yield of $1244\pm 114$ ($420\pm 38$) events with one (two or three) tag(s).

\begin{figure}[t!]
  \begin{center}
        \begin{tabular}{c}

	 \epsfig{file=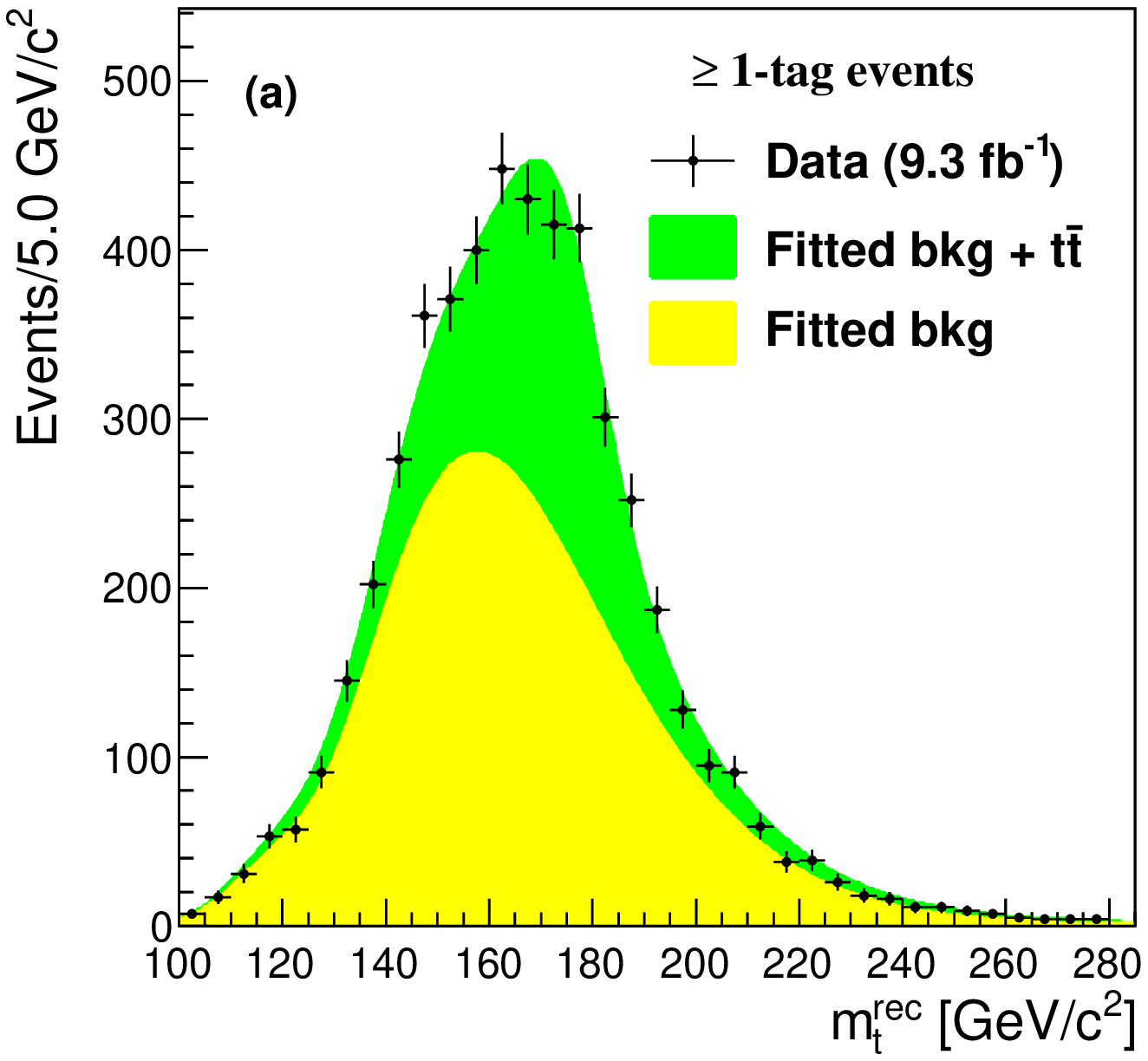,clip=true,bbllx=5pt,bblly=5pt,bburx=425pt,bbury=405pt,width=8cm} 
   
         \\   
	   
         \epsfig{file=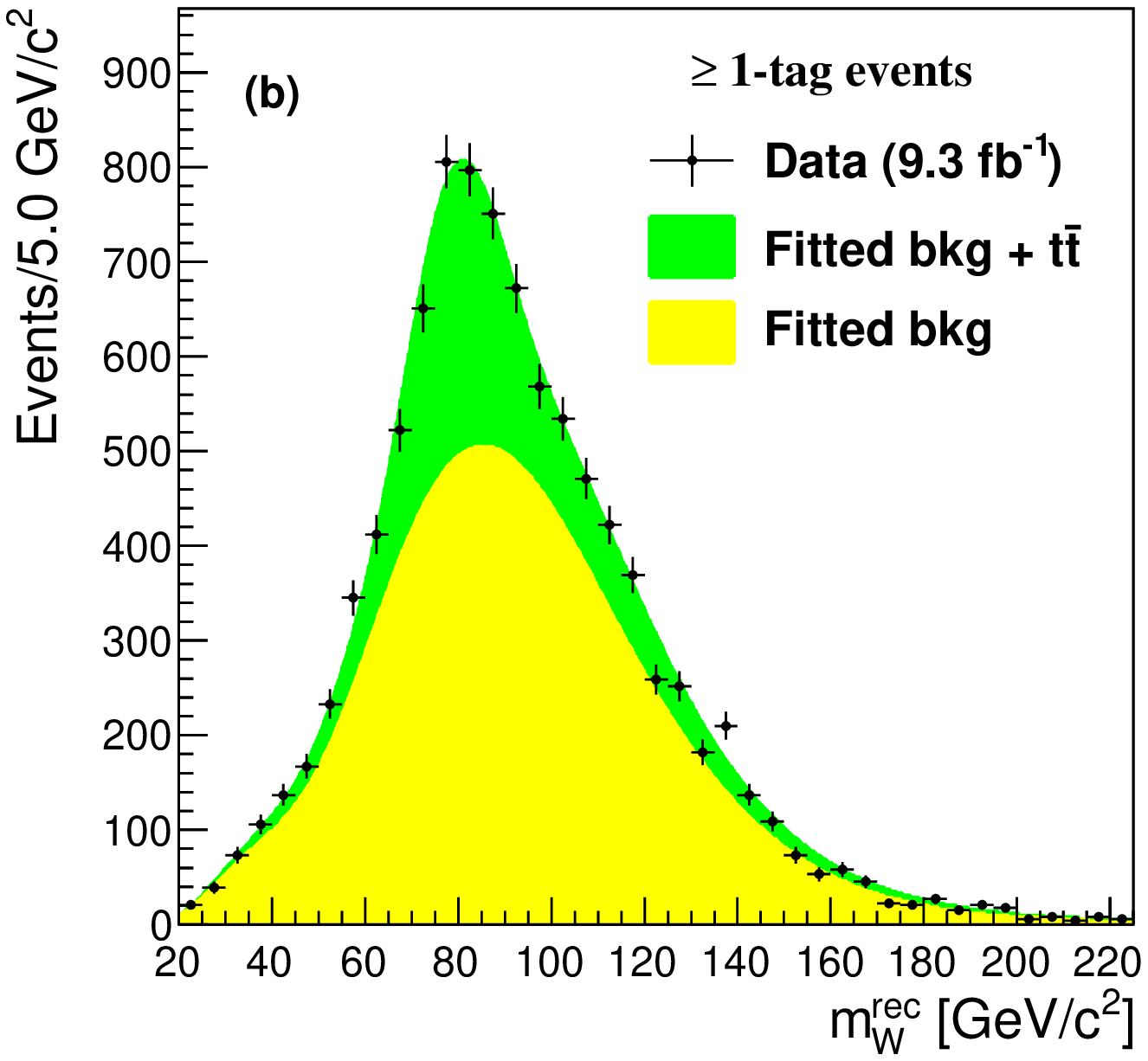,clip=true,bbllx=5pt,bblly=20pt,bburx=425pt,bbury=405pt,width=8cm}
   \end{tabular}
   \end{center} 
  \vspace*{-3mm}
   \caption{
            Distributions of $m_{t}^{\rm rec}$ (a) and $m_{W}^{\rm rec}$ (b) 
            for events  with one, two, or three tags (black dots),
            compared to the distributions from background and signal corresponding to the measured values of 
            $M_{\rm top}$ and $\Delta_{\rm JES}$. 
            The expected distributions are 
            normalized to the yields returned by the best fit. 
           }
   \label{fig:topmeas}
 \end{figure}

The distributions of 
 $m_{t}^{\rm rec}$ and $m_{W}^{\rm rec}$ 
for the data and the comparison with the expectation from the sum of background and signal for $M_{\rm top}$ 
and $\Delta_{\rm JES}$ corresponding to the measured values are shown in  Fig.\,\ref{fig:topmeas}. The contributions from  events with one, two, or three tags are summed together and the signal and background yields are normalized to the yields returned by the best fit. 

The measurements of  $M_{\rm top}$ and $\Delta_{\rm JES}$   are affected by various sources of systematic uncertainties, summarized in Table~\ref{tab:obssyst}.
These uncertainties  can be divided into four categories: (1) the modeling of signal events, including the choice of Monte Carlo generator and parton distribution function,
  the amount of initial and 
 final state radiation, and  the effects of color reconnections;
(2) the measurement method, including the dependence on the other free parameters of the fit, the size of the samples used to build the reference templates,
the variables used to perform calibration PEs like the $t\bar{t}$ production
cross section and the integrated luminosity of the data, and the trigger simulation;
(3) the background modeling, the $b$-tagging efficiency,
the effects of multiple hadron interactions (pileup) related to 
the instantaneous luminosity; and (4) the jet energy scale calibration.   
The largest contribution comes from the the jet energy scale calibration, 
given the large number of jets representing a typical feature of the
all-hadronic channel.
 With respect to Ref.\,\cite{ahprd}  
we  add in this analysis the uncertainties related to the background shape (and not to its normalization), to the $t\bar{t}$ cross section and to the integrated luminosity.
In general, the uncertainties are evaluated by performing PEs based on
templates made with specific variations of the original signal samples, taking
the  differences in the average values of $M_{\rm top}$ and $\Delta_{\rm JES}$ with respect
to the pseudoexperiments performed with default templates. Finally, 
possible residual biases remaining after the calibration and uncertainties on the parameters
of the calibration functions are taken into account. 
\begin{table}[tb!]
\caption{
          Sources of systematic uncertainties
          on the $M_{\rm top}$ and $\Delta_{\rm JES}$ measurements.
          The total uncertainty is evaluated as the quadrature sum of all contributions. 
         }
\begin{center}
  \begin{tabular}{lccc}
  \hline
  \hline

  \parbox[c][5mm][c]{6mm}{Source}  &~$\sigma_{M_{\rm top}}$~   & ~$\sigma_{\Delta_{ \rm JES}}$~ \\
  \parbox[c][5mm][c]{4mm}{      }  &     ~(GeV/$c^2$)~      &                                 \\
  \hline

  Generator (hadronization)               &  \parbox[c][4.5mm][c]{5mm}{$ 0.29 $}   & $ 0.273 $   \\ 
  Parton distribution functions           &  \parbox[c][6.5mm][c]{5mm}{ $ ^{+0.18}_{-0.36}$}   &  $ ^{+0.096}_{-0.052}$  \\
  Initial\,/\,Final state radiation       &  \parbox[c][4.5mm][c]{5mm}{$ 0.13 $}   &  $ 0.232 $  \\
  Color reconnection                      &  \parbox[c][4.5mm][c]{5mm}{$ 0.32 $ }   &  $ 0.101 $   \\

  $\Delta_{\rm JES}$ fit                      &  \parbox[c][4.5mm][c]{5mm}{$ 0.97 $}   & $  --   $   \\
  $M_{\rm top}$ fit                        &   \parbox[c][4.5mm][c]{5mm}{$ -- $}    & $  0.207$    \\
  Other free parameters of the fit        &  \parbox[c][4.5mm][c]{5mm}{$ 0.41 $}   & $  0.040   $   \\

  Templates sample size                 &  \parbox[c][4.5mm][c]{5mm}{ $ 0.34 $ }   &  $ 0.071 $  \\

  $t\bar{t}$ cross section                &  \parbox[c][4.5mm][c]{5mm}{ $ 0.15 $}   &  $ 0.034 $  \\ 
  Integrated luminosity                         &  \parbox[c][4.5mm][c]{5mm}{$ 0.15 $}   &   $ 0.032 $  \\
  Trigger                                 &  \parbox[c][4.5mm][c]{5mm}{ $ 0.61 $ }   & $ 0.188 $   \\ 

  Background shape                        &  \parbox[c][4.5mm][c]{5mm}{ $ 0.15 $ }   &  $ 0.014 $  \\ 
  $b$-tagging                             &  \parbox[c][4.5mm][c]{5mm}{$ 0.04 $}   & $ 0.018 $   \\ 
  $b$-jets energy scale                   &  \parbox[c][4.5mm][c]{5mm}{$ 0.20 $}   & $ 0.035 $    \\ 
  Pileup                                  &  \parbox[c][4.5mm][c]{5mm}{ $ 0.22 $}   &  $ 0 $  \\

  Residual JES                 &  \parbox[c][4.5mm][c]{5mm}{$ 0.57 $}   & $  --   $   \\

  Residual bias\,/\,Calibration          &  \parbox[c][6.5mm][c]{5mm}{$ ^{+0.27}_{-0.24}$}   &  $ ^{+0.077}_{-0.096}$  \\

  \hline

 Total                                   &  \parbox[c][6.5mm][c]{5mm}{$ ^{+1.55}_{-1.58} $}   & $  ^{+0.492}_{-0.488}$  \\

  \hline
  \hline
  \end{tabular}
\end{center}

\label{tab:obssyst}
\end{table} 

%\par
In summary, a measurement of the top-quark mass using top-quark pairs decaying into a fully hadronic final state is presented,
using $p\bar{p}$ collision data corresponding to the full integrated luminosity
of $9.3$\,fb$^{-1}$ collected by the CDF experiment in Run II.
The large background affecting this channel is strongly suppressed through
an optimized event selection, based on a neural network and the requirement of one, two, or three jets originating from $b$ quarks.
The simultaneous calibration of the jet energy scale allows us to reduce  the systematic uncertainty
due to this source to 0.97\,GeV/$c^2$. 
The measured value of the top-quark mass is 
$M_{\rm top}=175.07~\pm 1.19\,({\rm stat})~ ^{+1.55}_{-1.58}\,({\rm syst})$ GeV/$c^2$,
with a total uncertainty of approximately $2.0$\,GeV/$c^2$, corresponding to a 1.1\% relative uncertainty.
This final result in the all-hadronic channel is complementary to the most recent measurements obtained in 
other channels by the CDF Collaboration\,\cite{top-lepjets, top-metjets}, and consistent with the CMS measurement in the same channel\,\cite{cms_allhad}.
% 
% Acknowledgements

We thank the Fermilab staff and the technical staffs of the
participating institutions for their vital contributions. This work
was supported by the U.S. Department of Energy and National Science
Foundation; the Italian Istituto Nazionale di Fisica Nucleare; the
Ministry of Education, Culture, Sports, Science and Technology of
Japan; the Natural Sciences and Engineering Research Council of
Canada; the National Science Council of the Republic of China; the
Swiss National Science Foundation; the A.P. Sloan Foundation; the
Bundesministerium f\"ur Bildung und Forschung, Germany; the Korean
World Class University Program, the National Research Foundation of
Korea; the Science and Technology Facilities Council and the Royal
Society, United Kingdom; the Russian Foundation for Basic Research;
the Ministerio de Ciencia e Innovaci\'{o}n, and Programa
Consolider-Ingenio 2010, Spain; the Slovak R\&D Agency; the Academy
of Finland; the Australian Research Council (ARC); and the EU community
Marie Curie Fellowship Contract No. 302103.


\begin{thebibliography}{99}
\bibitem{coordinate} We use a cylindrical coordinate system where $\theta$ is the polar angle 
                     with respect to the
                     proton beam direction ($z$ axis), 
                     $\phi$ is the azimuthal angle about the beam axis, 
                     and the
                     pseudorapidity is defined as $\eta = - \ln \tan(\theta/2)$.
                     The transverse momentum, $p_{T}$, and tranverse energy,
                     $E_{T}$, of jets are given by $|p|\,\sin\theta$  and $E\, \sin\theta$  respectively.
                     The missing $E_{T}$ vector, $\vec{\met}$, is defined by 
                     $\vec{\met} = -\sum_{i}\,E_{T,i} \,\hat{n}_{T,i}$ where $\hat{n}_{T,i}$
                     is the unit vector in the $x$-$y$ plane pointing from the primary interaction
                     vertex to a given calorimeter tower $i$, 
                     and $E_{T,i}$ is the $E_{T}$ measured in that tower.
                     Finally $\met = |\vec{\met}|$. % Jan12
\end{thebibliography}
\end{document}